\documentclass[a4paper,12pt]{article}
\usepackage{graphics}
\usepackage{amsfonts}

\begin{document}
%

\newcommand{\be}{\begin{equation}}
\newcommand{\ee}{\end{equation}}
\newcommand{\bea}{\begin{eqnarray}}
\newcommand{\eea}{\end{eqnarray}}
\newcommand{\bean}{\begin{eqnarray*}}
\newcommand{\eean}{\end{eqnarray*}}
\font\upright=cmu10 scaled\magstep1
\font\sans=cmss12
\newcommand{\ssf}{\sans}
\newcommand{\stroke}{\vrule height8pt width0.4pt depth-0.1pt}
\newcommand{\Z}{\hbox{\upright\rlap{\ssf Z}\kern 2.7pt {\ssf Z}}}
\newcommand{\ZZ}{\Z\hskip -10pt \Z_2}
\newcommand{\C}{{\rlap{\upright\rlap{C}\kern 3.8pt\stroke}\phantom{C}}}
\newcommand{\R}{\hbox{\upright\rlap{I}\kern 1.7pt R}}
\newcommand{\HH}{\hbox{\upright\rlap{I}\kern 1.7pt H}}
\newcommand{\CP}{\hbox{\C{\upright\rlap{I}\kern 1.5pt P}}}
\newcommand{\identity}{{\upright\rlap{1}\kern 2.0pt 1}}
\newcommand{\half}{\frac{1}{2}}
\newcommand{\quart}{\frac{1}{4}}
\newcommand{\pr}{\partial}
\newcommand{\bm}{\boldmath}
\newcommand{\I}{{\cal I}} 
\newcommand{\M}{{\cal M}}
\newcommand{\N}{{\cal N}}
\newcommand{\e}{\varepsilon}

\thispagestyle{empty}
\rightline{DAMTP-2011-100}
\vskip 3em
\begin{center}
{{\bf \Large Monopole Planets and Galaxies
}} 
\\[15mm]

{\bf \large N.~S. Manton\footnote{email: N.S.Manton@damtp.cam.ac.uk}} \\[20pt]

\vskip 1em
{\it 
Department of Applied Mathematics and Theoretical Physics,\\
University of Cambridge, \\
Wilberforce Road, Cambridge CB3 0WA, U.K.}
\vspace{12mm}

\abstract{Spherical clusters of $SU(2)$ BPS monopoles are investigated 
here. A large class of monopole solutions is found using an abelian
approximation, where the clusters are spherically symmetric, although 
exact solutions cannot have this symmetry precisely. Monopole clusters 
generalise the Bolognesi magnetic bag solution of the same charge, but 
they are always larger. Selected density profiles give structures 
analogous to planets of uniform density, and galaxies with a density 
decaying as the inverse square of the distance from the centre. The 
Bolognesi bag itself has features analogous to a black hole, and this 
analogy between monopole clusters and astrophysical objects with or 
without black holes in their central region is developed further. It 
is also shown that certain exact, platonic monopoles of small charge 
have sizes and other features consistent with what is expected for 
magnetic bags.}

\end{center}

\vskip 3.5em
\leftline{Keywords: BPS Monopoles, Magnetic Bags, Planets, Galaxies,
Black Holes}
\leftline{PACS: 14.80.Hv, 11.27.+d, 96.12.Fe, 98.62.Ck, 04.70.Bw}
\vskip 1em

\vfill
\newpage
\setcounter{page}{1}
\renewcommand{\thefootnote}{\arabic{footnote}}


\section{Introduction} 
\vspace{4mm}

SU(2) Yang--Mills gauge theory with an adjoint scalar Higgs field is a classic
example of a nonlinear field theory in three space dimensions with smooth,
topological soliton solutions. The solitons are magnetic monopoles 
\cite{AH,book,Shn}. SU(2) magnetic monopoles were first discovered 
by 't Hooft and Polyakov, but the special case pinpointed by
Bogomolny, Prasad and Sommerfield (BPS) has received particular 
attention from physicists and mathematicians. Static BPS monopoles 
obey a first order partial differential equation known as 
the Bogomolny equation, ${\bf B}= {\bf D}\Phi$, where ${\bf B}$ is the 
Yang--Mills magnetic field and ${\bf D}\Phi$ is the
covariant gradient of the Higgs field $\Phi$. In addition, there is the
boundary condition that at spatial infinity, $\Phi$ is of unit
magnitude; in other words, $\Phi$ has unit vacuum expectation value. 
This Bogomolny equation is a three-dimensional analogue of the 
four-dimensional self-dual Yang--Mills equation, whose solutions are 
instantons. It is not easy to solve explicitly. 

Monopole solutions are characterised by a topological charge $N$, a
positive integer that can be identified with the number of basic
monopoles, so solutions with charge $N$ are $N$-monopole
configurations. The basic monopole of unit charge is spherically 
symmetric and has core size 1. Outside the core, the fields abelianize and 
resemble those of a Dirac monopole, with a magnetic Coulomb tail, 
and in addition a long-range Higgs field tail.

$N$-monopole solutions of the Bogomolny equation are plentiful 
\cite{JT}, even after the gauge freedom is quotiented out. The
magnetic and scalar forces between static BPS monopoles cancel 
\cite{Man}, so there is great freedom in where the constituent 
monopoles can be located, and it has been rigorously established 
that there is a connected $4N$-dimensional moduli space of 
$N$-monopole solutions. There is a deep mathematical theory, combining 
ideas of integrable systems, holomorphic bundles, Nahm data and 
spectral curves, and a mini-twistor space, giving insights into 
the Bogomolny equation and its monopole solutions \cite{AH,book}. 
The monopoles of charge $N$ are in 
1-1 correspondence with (based) rational maps of degree $N$, from 
$S^2$ to $S^2$, known as Jarvis maps in this context \cite{Jar}. By
considering Jarvis maps, it becomes clear that no BPS monopole of charge 
greater than $1$ can be exactly spherically symmetric, so a cluster of 
monopoles can be at most approximately spherically symmetric. 

The 1-monopole solution has a simple closed form, and much is known about 
2-monopole solutions \cite{AH}. A 2-monopole solution has just one interesting
parameter, the spatial separation of the pair. However, the
the field values and energy density are complicated functions of
spatial position and monopole separation. For charges higher than 2, 
only certain specially symmetric monopoles are known precisely 
\cite{book}. Even here, what is known are the spectral curves and the 
rational maps associated with the monopole. The fields and energy 
distributions need to be calculated numerically. Examples are a few monopoles 
with platonic symmetry, with charges $N \le 11$. There is
also a family of $N$-monopole solutions, for all $N$, which are 
regular polygons of $N$ unit charge monopoles \cite{HMM}. 
When the radius of the regular $N$-gon solution is minimised, the $N$ 
monopoles merge, and the energy density is concentrated in a circular 
ring, or torus \cite{FHP,PR,Hit}. The platonic monopoles, and the 
family of toroidal monopoles for all $N > 1$, are among the most 
compact monopole solutions. 

If all $N$ monopoles are very well separated from each other, then the 
solution is a superposition of $N$ particle-like objects, each with 
little distortion. The $4N$ parameters are the $N$ positions of the
unit charge monopoles ($3N$ parameters) and $N$ phases which are 
required to define the gluing of their fields into space. However, 
there is little insight so far into the spatial structure of a generic
$N$-monopole solution, where the monopoles form a relatively 
compact cluster. Monopoles cannot be very close together, as they 
get larger as they approach each other, so it is not clear how big 
such a cluster might be. 

The purpose of this paper is to shed some light on monopole clusters
of large charge, and our method is to extend the analysis that was 
used by Bolognesi to discover and investigate a further class of 
BPS monopole solutions called magnetic bags \cite{Bol1}. 

The existence of magnetic bags has not been rigorously proved, although it
is likely they exist for all large $N$. The bags are approximately 
spherical monopole solutions of charge $N$, with the nonabelian 
structure \cite{LW} concentrated on a thin spherical surface of radius 
approximately $N$. Because of the surface structure, the fields are
not exactly spherically symmetric. Bags can be distorted away from 
spherical, but we shall focus on those that are not. What is 
known precisely about magnetic bags are the abelianized gauge and 
Higgs fields that they produce. These resemble the electric field and 
the electrostatic potential arising from a charged conductor. 

The spherical magnetic bags are especially interesting, because, for
each $N$, they appear to be the most compact $N$-monopole solutions that
there are. We shall present new evidence for this in this paper. Our 
main purpose, however, is to generalise the construction of the
magnetic bags to a much larger class of solutions that can be
interpreted as spherical clusters of monopoles. Dilute clusters of 
$N$ unit charge monopoles are very large, because the size of each 
constituent monopole depends on the Higgs field in the neighbourhood. If
the background abelianized Higgs field is $\phi$, then locally a 
monopole has radius $1/\phi$. Our calculations below will give lower 
bounds on the size of monopole clusters based on this basic property of the 
constituent monopoles. In a spherical cluster, $\phi$ is smallest near 
the centre, so here the monopoles are largest. 

In Section 2 we shall review the properties of the Bolognesi magnetic bag, 
and then in Section 3 present a multi-layer generalisation of the magnetic 
bag, assuming spherical symmetry for simplicity. We shall establish an 
important constraint on the radial distribution of charge in such a structure. 
Various inequalities follow from this, and one of our conclusions will be 
that no multi-layer solution packs all $N$ monopoles into a smaller 
volume than the Bolognesi bag. In Section 4 we give a smoothed-out 
description of a multi-layer magnetic bag, which now becomes a 
cluster of monopoles with a quasi-smooth charge density. The density is a
rather arbitrary function of distance from the centre, subject to one
constraint. We discuss whether a Bolognesi type of bag
persists in the inner part of the cluster. 

In Sections 5 and 6 we present examples of monopole clusters that have 
density profiles typical of two types of astrophysical object -- the first 
is a planet, with a constant density inside and zero density outside; the 
second is a spherical galaxy, with a density falling with the inverse 
square of the distance from the centre \cite{BT}. We give these two 
types of cluster the names monopole planet and monopole galaxy.
There are some analogies between monopole clusters subject to
Yang--Mills--Higgs forces, and astrophysical objects subject to
gravitational forces, including black holes. We shall not 
simultaneously consider both types of force, although Bolognesi 
has recently considered gravitating magnetic bags \cite{Bol2}.
  
In Section 7, which depends on more technical material about exact 
monopole solutions, we show, by calculating spectral radii, that the 
known monopoles with platonic symmetries are examples of approximately 
spherical magnetic bags, of small charge. We also show that the 
toroidal monopole of charge $N$ is not as compact as the spherical 
magnetic bag of charge $N$. 

In Section 8 we present our conclusions, and summarise the analogy
that appears to exist between monopole clusters and astrophysical
objects, which extends the analogy between magnetic bags and 
black holes that has been noticed by Bolognesi \cite{Bol2}. 

\vspace{7mm}
\vfill
\newpage

\section{Magnetic Bags}
\vspace{4mm}

An interesting structure, discovered by Bolognesi and called a magnetic 
bag \cite{Bol1}, is possible for a monopole with a large charge $N$. It is 
an abelian approximation to what is expected to be an exact solution of the
$SU(2)$ Bogomolny equation ${\bf B}= {\bf D}\Phi$. The simplest
version of the bag is spherically symmetric. The abelian fields are a
scalar field $\phi$, representing the magnitude of the Higgs field
$\Phi$, and a magnetic field ${\bf b}$, the projection of the $SU(2)$
magnetic field ${\bf B}$
onto the direction of $\Phi$ in the $SU(2)$ Lie algebra. They are
related by ${\bf b} = \nabla \phi$, which is the abelianized Bogomolny
equation. Away from the monopoles $\nabla \cdot {\bf b} = 0$,
so $\nabla^2\phi = 0$; the monopoles are Coulomb-type sources for ${\bf b}$. 
In the interior of the bag $\phi=0$,
and in the exterior $\phi = 1 - N/r$. Correspondingly, ${\bf b}$
vanishes in the interior and ${\bf b}= N{\bf r}/r^3$ in the
exterior. At the bag surface, separating interior from exterior,
$\phi$ is required to be continuous, so the bag is at radius $R =
N$. Note that $\nabla\phi$ and ${\bf b}$ are discontinuous across the
bag, the discontinuity of the flux of ${\bf b}$ being $4\pi N$. The
interpretation is that the bag carries magnetic charge $N$,
spread uniformly over the surface, and that the interior of the bag is
a magnetic conductor, carrying no charge.

It is very plausible that there is an exact solution, approximated 
by the bag, which has abelianized fields as
described above except in a thin spherical wall at radius $N$, where
the fields are fully nonabelian. The bag wall is of thickness
approximately $\sqrt{N}$, and is a slightly curved version of the flat
monopole wall solution found by Ward \cite{War}. That solution has a
two-dimensional lattice periodicity, which is why the bag cannot be
exactly spherically symmetric. The spherical bag must not only curve
the wall, but also introduce defects in the two-dimensional lattice 
structure. Nevertheless, the lattice structure probably survives 
locally. Lee and Weinberg have described in considerable detail some 
of the topological features of the nonabelian fields of the bag \cite{LW}.
Nahm data associated with magnetic bags of large charge have been
discussed by Harland \cite{Har}.

Some evidence for the existence of bags comes from the platonic
monopoles \cite{HMM,HS1,HS2}. These are exact solutions of the Bogomolny 
equation of low charge which resemble bags, in that they are hollow 
platonic polyhedra, with energy and nonabelian field structures 
concentrated on their surfaces. So far, their radii have not been
compared with the Bolognesi estimate $R=N$. Using the spectral
curves of these monopoles, we calculate their radii in Section 7 
here, and show that they are close to the Bolognesi value. As a 
definition of size we use the spectral radius \cite{AH}. 

Perhaps the most surprising property of a magnetic bag is its large
size. Since the radius is $N$, its area is a multiple of $N^2$ and its
volume a multiple of $N^3$. One might expect BPS monopoles to be a model
of rather normal matter, where $N$ particles fit into a volume of
order $N$, but this is not the case. The magnetic bag is much larger. 
Black holes have a similarly exotic size scaling. The radius of the
horizon of a Schwarzschild black hole of mass $M$ is $2M$ in natural 
units, and the black hole area is proportional to $M^2$. A magnetic
bag is therefore, to some extent, a gauge theory analogue of a black hole. 

It is worth recalling a few more properties
of the magnetic bag, and some variants of the bag, pointed out by
Bolognesi, and by Lee and Weinberg. First, the bag can be deformed
from spherical, because of the large number of dimensions of the
moduli space of $N$ monopoles. However, we will only consider spherical 
distributions of monopoles here. Next, there is a rather trivial rescaling 
of the Bolognesi bag if the vacuum expectation value of the Higgs field
$\Phi$ is $v$, rather than the special value $v=1$. One now has $\phi
= v - N/r$ outside the bag, and $\phi = 0$ inside, so the bag radius
is $R = N/v$. Its thickness is $\sqrt{N}/v$ \cite{LW}. The argument for this is
that the bag area is $N^2/v^2$ (dropping constants of order 1), so the
area per unit charge monopole is $N/v^2$. The characteristic length
associated with a monopole in the bag surface (which can be thought of
as the monopole separation within the surface) is therefore the square
root of this, $\sqrt{N}/v$, and since it is the only length scale in
the bag, the bag thickness must be the same. We will, in fact, always 
suppose that $v=1$ and impose the boundary condition that $\phi \to 1$ 
as $r \to \infty$. However, the bag solution with other values of $v$ 
will be locally relevant. Note that for any $v$ the bag thickness is
$R/\sqrt{N}$, where $R$ is the radius and $N$ the charge. 
 
Lastly, we recall that there is a variant bag of charge $N$ where
$\phi$ does not vanish inside, but instead $\phi = c$ inside, with
$0<c<1$. Outside, $\phi = 1 - N/r$ as usual, so the bag radius is
$R = N/(1-c)$, making $\phi$ continuous across the bag surface.  This bag
radius is larger than the radius of the Bolognesi bag, and if $c$ is
not very small, then the bag breaks up into constituent, unit
charge monopoles. To estimate the critical value of $c$ where break-up
occurs, note that since the Higgs field is $\phi = c$ in the bag vicinity, unit
charge monopoles have a radius $1/c$. The bag has broken up into
constituents if the monopole separation within the bag, $\sqrt{N}$, 
is bigger than this. The critical value of $c$ is therefore 
$1/\sqrt{N}$, which is much less than 1 if $N$ is large. The critical 
value of $c$ is reached when the bag has expanded to a radius 
approximately $N + \sqrt{N}$. If the bag is larger than this it should 
be regarded as a spherical layer of separated, unit charge monopoles, 
rather than as a Bolognesi bag.

\vspace{7mm}

\section{Multi-layer Magnetic Bags}
\vspace{4mm}

Using the abelianized Bogomolny equation ${\bf b} = \nabla \phi$, 
it is straightforward to
calculate the Higgs field $\phi$ and magnetic field ${\bf b}$ of a
multi-layer structure of several magnetic bags with spherical symmetry. Let us
suppose that there are bags with positive integer charges $\{ N_1,
N_2, \dots, N_K\}$ on spheres of radii $\{ R_1, R_2, \dots, R_K\}$,
where $0 < R_1 < R_2<\dots<R_K$. The total charge is $N=N_1+N_2+\cdots
+N_K$. Let $b$ denote the magnitude of the radial magnetic field
${\bf b}$. $b(r)$ has discontinuities at each bag. Between $R_k$ and
$R_{k+1}$, 
\be 
b(r) = \frac{N_1 + N_2 + \cdots + N_k}{r^2} \,,
\ee
the magnetic analogue of the Coulomb field due to the enclosed charge.
Also $b(r) = 0$ for $r<R_1$ and $b(r)=N/r^2$ for $r>R_K$. The Higgs field
$\phi(r)$ satisfies
\be
\frac{d\phi}{dr} = b(r) \,.
\ee
If we integrate out from $r=0$, and assume $\phi(0)=0$, then we find
that $\phi(r)=0$ for $r<R_1$,
\be
\phi(r) = \frac{N_1}{R_1} + \frac{N_2}{R_2} + \cdots + \frac{N_k}{R_k} -
\frac{N_1 + N_2 + \cdots + N_k}{r}
\ee
between $R_k$ and $R_{k+1}$, and
\be
\phi(r) = \frac{N_1}{R_1} +\frac{N_2}{R_2}+\cdots + \frac{N_K}{R_K} -
\frac{N}{r}
\ee
for $r>R_K$. $\phi$ is continuous at each bag radius, but
$d\phi/dr$ has a discontinuity.

The boundary condition $\phi \to 1$ as $r \to \infty$ requires that
\be
\label{ConstraintDiscrete}
\frac{N_1}{R_1} + \frac{N_2}{R_2} + \cdots + \frac{N_K}{R_K} = 1 \,.
\ee
This is the key constraint on multi-layer bags and their distribution
of charges. This constraint is the generalisation of the result that
for a single Bolognesi bag, $R=N$. Various inequalities follow. Among them we
have
\be
\frac{N_k}{R_k} < 1 \,,
\ee
so each bag is larger than it would be if it had the same charge but
were alone. More interesting is to consider the total charge
$N$. Replacing each radius $R_k$ by $R_1$ we increase the sum on the
left hand side of eq.(\ref{ConstraintDiscrete}) to a value greater
than 1. But the sum is now $N/R_1$, so $R_1<N$. Similarly, replacing
$R_k$ by $R_K$, we find $R_K>N$. Thus, compared with a Bolognesi bag
of charge $N$, the multi-layer structure has a smaller innermost bag,
and a larger outermost bag. The structure as a whole is larger than 
the Bolognesi bag. Indeed, this is strong evidence that the smallest 
radius into which one can fit $N$ monopoles is the Bolognesi radius $R=N$.
 
A very similar solution with $\phi(r) = c$ for $r<R_1$ (with $0<c<1$)
exists when
\be
\label{ConstraintDiscretec}
\frac{N_1}{R_1} + \frac{N_2}{R_2} + \cdots + \frac{N_K}{R_K} = 1-c \,.
\ee
Here $R_K>N/(1-c)$, so the solution is larger than before.

Let us now consider the type of structure each bag layer has. For the 
rest of this section we assume $c=0$. The innermost layer is a bag of 
Bolognesi type. It has charge $N_1$ and radius $R_1$, the Higgs field 
is $\phi = 0$ inside and $\phi(r) = N_1/R_1 - N_1/r$ outside (as far 
as the second layer). This is just as expected for a Bolognesi bag of 
charge $N_1$ and effective vacuum expectation value $v=N_1/R_1$. The
thickness of this bag is $\sqrt{N_1}/v = R_1/\sqrt{N_1}$, which is small
compared to $R_1$ provided $N_1$ is large.  The second layer needs to
be further from the first than this thickness, for the multi-layer
point of view to be justified.
 
Let us therefore assume that the separation $R_2 - R_1$ is significantly 
larger than $R_1/\sqrt{N_1}$, but significantly less than $R_1$. Let us 
also assume that the charges in the first and second layer are
similar. Within the second layer the monopole separation is 
$R_2/\sqrt{N_2}$, which is comparable to the separation within the first 
layer. Just inside the second layer the Higgs field is 
\be
\phi = \frac{N_1}{R_1} - \frac{N_1}{R_2} 
= \frac{N_1(R_2-R_1)}{R_1R_2}
\ee
which, by our assumptions, is considerably larger than both $\sqrt{N_1}/R_2$ 
and $\sqrt{N_2}/R_2$. We estimate the monopole size in the second layer 
as $1/\phi = R_1R_2/(N_1(R_2-R_1))$, and this is considerably less than 
the monopole separation within the layer, $R_2/\sqrt{N_2}$. The second layer 
is therefore not a Bolognesi bag, but instead a spherical arrangement 
of well separated unit charge monopoles. Since $\phi$ increases as one 
moves out to subsequent layers, the characteristic size of a monopole 
decreases. On the other hand, the monopole separation within each
layer increases as one moves out, provided the charges in all the 
layers are similar. It follows that all the outer layers 
consist of well separated monopoles, rather than a bag of Bolognesi 
type. The entire solution consists of an inner layer which is a 
Bolognesi bag, surrounded by a layered cluster of well separated monopoles. 
These monopoles are not of equal size. They are largest near the
centre, where the Higgs field is small, and decrease in size further 
out, with monopoles in the outer layers having approximately unit size.

A multi-layer bag solution may have a Bolognesi type
of bag as one of the intermediate layers. But this only happens if the
distribution of charges is very uneven, with the constraint
(\ref{ConstraintDiscrete}) satisfied by having one ratio $N_k/R_k$
very close to 1, and all the others much less than 1. The $k$th layer
is then a Bolognesi bag. It is still possible for one of the layers
closer to the centre to be a Bolognesi bag too, but its charge must be
very much smaller. We can verify this by considering a
structure with just two layers, each with a large charge (so that the
square root of the charge is small compared with the charge itself). 
If we fix their radii at $R_1$ and $R_2$ (with ratio of order 1), and
satisfy the constraint
\be
\frac{N_1}{R_1} + \frac{N_2}{R_2} = 1
\ee
by setting $N_1/R_1 = \varepsilon$ and $N_2/R_2 = 1 - \varepsilon$
with $\varepsilon$ small, then the inner layer is a Bolognesi bag
automatically, and the outer layer is one too, provided $\phi$ is
sufficiently small at $r=R_2$. This occurs if $\varepsilon <
1/\sqrt{N_2}$ . Since $N_2$ is approximately $R_2$, and $N_1 =
\varepsilon R_1 < \varepsilon R_2 \simeq \varepsilon N_2$, $N_1$ needs
to be less than $\sqrt{N_2}$. As anticipated, the inner charge is much
smaller than the outer charge.

Our main conclusion is that multi-layer bag structures are possible,
but the layers are not generally of the Bolognesi bag type. Typically,
the innermost layer is a Bolognesi bag, and it is surrounded by spherical
layers of isolated unit charge monopoles. We shall next consider the
continuum limit of these structures. There will be a spherically
symmetric density of magnetic charge located between an inner and outer
radius. At the inner radius we shall find there is often a Bolognesi
bag, and outside this, a gas of isolated monopoles.

\vspace{7mm}

\section{A Smooth, Spherical Cluster of Monopoles}
\vspace{4mm}

For a spherical cluster of BPS monopoles, of very large charge, the 
continuum limit arises when the layers discussed above merge into
a charge density that is a quasi-continuous function of the radius. 
It is convenient to work with the radial density $q(r)$, defined so 
that the charge in the shell between radius $r$ and radius $r+dr$ is 
$q(r)dr$. For monopoles satisfying the Bogomolny equation, $q(r) \ge 0$
everywhere. The charge $Q(R)$ interior to radius $R$ is
\be
Q(R) = \int_0^R q(r) \, dr \,,
\ee
and the magnetic field strength is $b(R) = Q(R)/R^2$. We shall assume 
that the total charge, $Q = {\rm lim}_{R \to \infty}Q(R) = N$, is finite.

The constraint (\ref{ConstraintDiscrete}) becomes, in the continuum limit,
\be
\int_0^\infty \frac{q(r)}{r} \, dr = 1 \,.
\ee
This is in the case that the Higgs field $\phi$ runs from $0$ at the origin
to $1$ at infinity. More generally, if $\phi$ has a positive value $c$ 
at the origin,
\be
\label{ConstraintContc}
\int_0^\infty \frac{q(r)}{r} \, dr = 1-c \,.
\ee
These integrals automatically converge at infinity if the total
charge is finite, but will fail to converge at the origin unless the
density $q$ tends to zero sufficiently fast there. Having a finite,
non-zero volume density of charge at the origin makes $q(r)$ proportional 
to $r^2$ for small $r$, and this is one way to ensure convergence. 
We shall also consider $q$ strictly vanishing inside an interior radius 
$R_{\rm in}$. It also makes sense to assume that $q$ vanishes outside some
outer radius $R_{\rm out}$. This is because monopoles each have at
least one unit of charge, so that a smoothly decaying charge density
stretching to infinity is not possible.

With the outer radius assumed, the total charge is
\be
Q= \int_0^{R_{\rm out}} q(r) \, dr \,,
\ee
and the constraint (\ref{ConstraintContc}) is
\be
\int_0^{R_{\rm out}} \frac{q(r)}{r} \, dr = 1-c \,.
\label{ConstraintRout}
\ee
Replacing $r$ by $R_{\rm out}$ in the denominator gives the inequality
$Q/R_{\rm out} < 1-c$ (assuming $q$ not entirely concentrated at 
$R_{\rm out}$), so the ball that contains all the charge has radius 
larger than $Q/(1-c)$, the radius of the one-layer bag for the same 
total charge.

We will need the formula for the Higgs field $\phi$. By analogy with
the multi-layer case, or by direct integration, we find
\be
\phi(R) = \int_0^R \frac{q(r)}{r} \, dr - \frac{Q(R)}{R} + c \,.
\ee
This is analogous to the Newtonian gravitational potential for a spherical mass
distribution, shifted so that $\phi = 1$ at infinity.
Using the constraint (\ref{ConstraintRout}), we have the alternative
expression
\be
\label{HiggsR}
\phi(R) = 1 - \int_R^{R_{\rm out}} \frac{q(r)}{r} \, dr - \frac{Q(R)}{R} \,.
\ee
$\phi(R)$ obeys the spherically symmetric version of the Poisson equation
\be
\frac{d}{dR}\left(R^2 \frac{d\phi}{dR}\right) = q(R) \,,
\ee
and $b(R) = d\phi/dR$ as before.

For generic densities $q(r)$, we anticipate that if $c=0$ a Bolognesi
bag could occur at some inner radius, with a gas of isolated monopoles
outside. For sufficiently non-uniform $q(r)$ there could be Bolognesi
bags at different radii, as we saw with the multi-layer structures,
but we will not seek these out again. In the next sections, we
consider two specific densities, corresponding to what we think of as
a monopole planet and a monopole galaxy. The monopole galaxy has a Bolognesi
bag at its inner radius.

\vspace{7mm}

\section{Monopole Planets}
\vspace{4mm}

A (small) planet is a spherical body where gravity plays a limited role.
If the material is uniform,
the density is uniform, dropping suddenly to zero at the planet's
outer radius. By a monopole planet, we mean a spherical cluster of
monopoles with a radial density $q(r) = 4\pi \rho r^2$ between $r=0$
and $r=R_{\rm out}$, with $\rho$, the charge 
density per unit volume, a positive constant. The density vanishes
outside. As the volume per unit charge is $\rho^{-1}$ we estimate the 
nearest neighbour monopole separation as $\rho^{-1/3}$. Here, and in
the rest of this and the next section, factors of order 1 are often ignored.

The total charge of the planet is 
\be
Q= \frac{4\pi}{3} \rho  R_{\rm out}^3 \,,
\ee
and the constraint (\ref{ConstraintRout}) simplifies here to
\be
2\pi \rho  R_{\rm out}^2 = 1-c \,.
\ee
Eliminating $\rho$ in favour of $Q$ we find
\be
\label{Rout}
R_{\rm out} = \frac{3}{2(1-c)} Q \,.
\ee
This is always greater than the Bolognesi radius $R=Q$.

One might imagine that for given charge $Q$, $R_{\rm out}$ is smallest
when $c=0$, but this choice of $c$ is not
consistent. Close to the origin $\phi$ would be so small that the
monopole sizes would exceed their separation. To restore consistency
we would need to dilute the monopole density near the origin, creating
a hole. We do not want to do this here, but will do so in the next
section. Instead, let us keep the density uniform, but find the minimal
consistent value of $c$. This is the value $c = \rho^{1/3}$, for which 
the monopole size close to the origin, which is $1/c$, matches the 
monopole separation $\rho^{-1/3}$. Again we can eliminate $\rho$ 
in favour of $Q$. We find that $c = O(Q^{-2/3})$ if $Q$ is large. So 
the monopoles close to the centre of the planet have size of order 
$Q^{2/3}$ (and hence volume $Q^2$), much larger than the monopoles
near the outside, which have unit size. From (\ref{Rout}) it follows 
that this densest possible monopole planet has size 
$R_{\rm out} = 3Q/2 + O(Q^{1/3})$. The Higgs field $\phi$ is small 
near the centre but not zero, so there is no magnetic bag in the
centre. A less dense monopole planet of the same charge arises if $c$
has a larger value.

\vspace{7mm}

\section{Monopole Galaxies}
\vspace{4mm}

The remarkable feature of many spherical and almost spherical
galaxies is that their density is highly non-uniform, decaying as the
inverse square of distance from the centre \cite{BT}. This density
profile is called a singular isothermal sphere. We shall investigate here 
the structure of a BPS monopole of large charge with such a density profile. 
If the charge density per unit volume decays as $r^{-2}$ then the radial
density $q(r)$ is a constant, $q_0$. Let us suppose that $q(r) = q_0$
in the radial range $R_{\rm in} \le r \le R_{\rm out}$ and vanishes
outside this range. The total charge $Q$ is $q_0(R_{\rm out}
-R_{\rm in})$. We want a finite charge, so $R_{\rm out}$ is finite. 
If $R_{\rm in} \ll R_{\rm out}$ then the total charge is approximately 
$q_0 R_{\rm out}$.

The usual constraint on the radial density becomes
\be
\int_{R_{\rm in}}^{R_{\rm out}}\frac{q_0}{r} \, dr = 1-c \,,
\ee
that is, 
\be
q_0 (\log R_{\rm out} - \log R_{\rm in}) = 1-c \,.
\ee
This constraint forces $R_{\rm in}$ to have a positive value. Having 
an inner cut-off at a positive radius is the same as in galaxy
modelling, and avoids a singular density at the origin. The
interesting case is where $c=0$. The Higgs field $\phi$ then vanishes 
for $r \le R_{\rm in}$, and we anticipate that there can be a Bolognesi bag
in the neighbourhood of $R_{\rm in}$, surrounded by a more dilute gas
of well separated monopoles. Note that $q_0$ is not very different
from 1 for a large range of ratios $R_{\rm out}/R_{\rm in}$.

We need the Higgs field $\phi$ between $R_{\rm in}$ and $R_{\rm out}$. The
expression (\ref{HiggsR}) reduces to
\be
\phi(R) = 1 - q_0(\log R_{\rm out} - \log R) - 
q_0\left(1 - \frac{R_{\rm in}}{R}\right) \,.
\ee
For $R$ near $R_{\rm out}$ this has the characteristic logarithmic
dependence on $R$ familiar from the Newtonian gravitational potential
in galaxy models. Near $R_{\rm in}$ there is a quadratic dependence 
on $R - R_{\rm in}$, and the slope of $\phi$ (to leading order) is
\be
\frac{d\phi}{dR} = \frac{q_0}{R_{\rm in}^2}(R-R_{\rm in}) \,.
\label{slope}
\ee

Now recall that for the multi-layer structures discussed in section 3,
with $\phi = 0$ in the interior, the innermost layer is usually a
Bolognesi bag. The charge $N_1$ and radius $R_1$ are arbitrary, but
they are related to the slope of $\phi$ just outside the bag, which is
$N_1/R_1^2$. The continuum distribution of charge density also allows
a Bolognesi bag by this criterion, as the following argument shows. 
Suppose the bag has thickness $l$,
small compared with $R_{\rm in}$, so the bag lies between $R_{\rm in}$
and $R_{\rm in} + l$. The charge in the bag is $q_0 l$ and the bag
radius is $R_{\rm in}$. The expected slope of the Higgs field just
outside the bag is $q_0 l/R_{\rm in}^2$. But this agrees with the
slope (\ref{slope}). So a Bolognesi bag is present, but this
calculation doesn't determine its thickness, nor its charge. However
there is a further consideration, as we have specified the charge
density throughout the monopole galaxy, and in particular, near the
inner radius. The volume per unit charge here is $ R_{\rm in}^2/q_0$
(dropping a $4\pi$ factor), so the separation of monopoles is
\be 
s = \frac{R_{\rm in}^{2/3}}{q_0^{1/3}} \,.
\ee
If we identify $s$ with $R_{\rm in}/\sqrt{q_0 l}$, the separation of 
monopoles within a Bolognesi bag of radius $R_{\rm in}$ and charge $q_0 l$, 
we deduce that $l=R_{\rm in}^{2/3}/q_0^{1/3}=s$. The thickness of the 
bag $l$ is the same as the monopole separation within the bag, 
which is what we expect. The charge of the bag is 
$Q_{\rm bag} = (q_0 R_{\rm in})^{2/3}$.

So we have now consistently determined the parameters of the Bolognesi 
bag at the centre of a monopole galaxy. In terms of the total 
charge $Q$ and the inner and outer radii of the monopole galaxy, the 
bag charge is
\be
Q_{\rm bag} = Q^{2/3} \left(\frac{R_{\rm in}}{R_{\rm out}}\right)^{2/3} \,,
\ee
using the approximation that $Q = q_0 R_{\rm out}$. $Q_{\rm bag}$ is
very much smaller than $Q$, so almost all monopoles form a dilute gas 
outside the bag. Similarly, the thickness is
\be
l =  \frac{R_{\rm in}^{2/3} R_{\rm out}^{1/3}}{Q^{1/3}} \,.
\ee

Therefore, a monopole galaxy has a feature rather analogous to
the black hole at the centre of a real galaxy, namely an inner
magnetic bag where monopoles have coalesced, carrying a small fraction 
of the total charge. Given the bag radius, and the outer radius and 
total charge of the monopole galaxy, the charge of the bag is 
determined. This contrasts with the black hole in a galaxy, whose mass 
is not theoretically determined.

We can have some fun with the monopole arithmetic. Suppose a
monopole galaxy has charge $10^{12}$, and the ratio of outer to inner
radius is $10^3$. Then the magnetic bag at the centre has charge $10^6$. These
numbers resemble the analogous numbers for a real galaxy and its
black hole (with charge identified with number of solar masses). So
the monopole equations, which forbid $\phi$ to be negative,
provide a useful model for regularising the density singularity inside
a galaxy, as well as providing an interesting analogy for
understanding why central black holes occur in galaxies.
   
\vspace{7mm}

\section{Spectral Radii of Monopoles}
\vspace{4mm}

This section depends on the theory of monopole spectral curves, and
the details, other than the next paragraph, may be skipped.

There are BPS monopole solutions of charges $N=3,4,5,7$ with the symmetries 
and shapes of the platonic solids. They provide small scale models of 
the Bolognesi bag that is expected at large $N$. Here we use the known
spectral curve of each of these monopoles to calculate its spectral 
radius $D_N$, which is a precisely defined estimate of the monopole 
size, and compare with the Bolognesi estimate of the radius,
$R_N$. We will show them to be similar. We include the $N=1$ and toroidal 
$N=2$ monopoles in the calculations, and also make some remarks about 
the sequence of axisymmetric, toroidal monopoles that exist for all $N$.
 
Any BPS monopole is characterised by its spectral curve. This is
obtained easily from the Nahm data of the monopole, or equivalently,
using the normalisable solutions of the Hitchin equation in
the monopole background. For details see, e.g. \cite{AH,book}. The 
spectral curve is a
complex curve lying in the space of all straight lines in $\R^3$ (which
is the mini-twistor space $TP^1$, a non-compact complex surface). The
complex coordinates of $TP^1$ are $\xi$ parametrising the direction 
of a straight line (obtained by usual stereographic projection), and 
$\eta$ which parametrises the intersection point of the
line with a (complex) plane orthogonal to the direction $\xi$. Given a
line with coordinates $\xi$ and $\eta$, the points 
$(x_1,x_2,x_3) \in \R^3$ that lie on it are those satisfying
\be
\eta - (x_2 - ix_1) + 2x_3 \xi + (x_2 +ix_1)\xi^2 = 0 \,.  
\ee
Important for us is to note that the distance between the line 
and the origin in $\R^3$ (the distance to the closest point on the line) is
\be
d = \frac{|\eta|}{1 + |\xi|^2} \,.
\ee
If $\eta = 0$ then the line passes through the origin. 

The spectral curve of a given monopole of charge $N$ is an algebraic
curve in $TP^1$ of the form 
\be
P_N(\eta,\xi)=0 \,,
\ee
where $P_N$ is a polynomial of degree $N$ in $\eta$ and (generically) 
of degree $2N$ in $\xi$. 
For fixed direction $\xi$, there are $N$ solutions $\eta$. So the
monopole has $N$ straight lines associated with it in this
direction (and in every other direction). These are called spectral
lines. Heuristically, spectral lines pass through regions where the 
energy density of the monopole is concentrated and the
fields are fully non-abelian. In particular, for $N$ well separated unit 
charge monopoles, these lines pass through all the monopole cores.

The spectral curve is compact. This implies that the
distance $d$ defined above has a finite maximum value $D$ as $\xi$ and
$\eta$ vary over the spectral curve. Let us call $D$ the spectral
radius of the monopole. There is a notion of a centred monopole (with centre
of mass at the origin). The spectral radius is most useful when applied to
a centred monopole, and is invariant under rotations of the monopole.

Let us now evaluate the spectral radii of the most symmetric, 
centred monopoles of low charge. We consider the basic $N=1$, 
$O(3)$-symmetric monopole, the toroidal $N=2$ monopole, and the bag-like 
monopoles of charges $N=3,4,5,7$ with tetrahedral, cubic, octahedral 
and dodecahedral form \cite{HMM,HS1,HS2}. There is also a monopole of 
charge $N=11$ with icosahedral form \cite{HMS}. They are unique for these 
charges, and no monopoles of smaller charge have these platonic 
symmetries. 

The spectral curves of all these have been determined, 
except the $N=11$ example. They are, respectively\footnote{Included 
in equations (\ref{spec4cube}) and (\ref{spec5octa}) is a factor 16 
correction to equations in \cite{HMM,HS2}, see \cite{Sut}.},
\bea
\eta &=& 0 \,,
\label{spec1hedge} \\
\eta^2 + \frac{\pi^2}{4} \xi^2 &=& 0 \,,
\label{spec2ax} \\
\eta^3 
+ \frac{\Gamma \left(\frac{1}{6}\right)^3\Gamma \left(\frac{1}{3}\right)^3}
{48\sqrt{3}\pi^{3/2}} i\xi(\xi^4 - 1) &=& 0 \,, 
\label{spec3tetra} \\
\eta^4 +\frac{3 \Gamma \left(\quart\right)^8}{1024\pi^2} 
(\xi^8 + 14 \xi^4 +1) &=& 0 \,, 
\label{spec4cube} \\
\eta^5 +\frac{3\Gamma \left(\quart\right)^8}{256\pi^2} 
(\xi^8 + 14 \xi^4 +1)\eta &=& 0 \,,
\label{spec5octa} \\
\eta^7 +  \frac{\Gamma \left(\frac{1}{6}\right)^6\Gamma
\left(\frac{1}{3}\right)^6}{64\pi^{3}} \ \xi(\xi^{10} + 11\xi^5-1)\eta
&=& 0 \,.
\label{spec7dodec}
\eea
Use of the identity $\Gamma(1/6) = (3/\pi)^{1/2}2^{-1/3}(\Gamma(1/3))^2$
gives alternative expressions for the constants above \cite{HMR}. 
In the cases $N>2$, the polynomial ${\cal K}(\xi)$ that appears here 
is a Klein polynomial associated with the relevant platonic solid \cite{Kle}. 
This is a polynomial that vanishes in the directions of all the face 
centres, or all the edge centres, or all the vertices of the solid.  
Note that we are following the conventions of \cite{book} in this 
section, where the asymptotic behaviour of the Higgs field is 
$\phi = 1 - N/2r$. The Bolognesi radius is therefore $R_N = N/2$.

For given $\xi$ it is easy to find the roots $\eta$ for each
of these equations. One of the roots is $\eta=0$ in cases (\ref{spec5octa}) 
and (\ref{spec7dodec}). The non-zero roots all have the same magnitude 
$|\eta|_{\xi}$. Evaluating $d_{\xi}=|\eta|_{\xi}/(1 + |\xi|^2)$ 
therefore gives the maximum distance from the origin of spectral lines 
in the direction $\xi$. $d_{\xi}$ is a function with the symmetry of 
the monopole. The spectral radius, which we denote by $D_N$ for these special 
monopoles, is the maximum value of $d_{\xi}$ over all values of $\xi$.

For $N=1$, $d_{\xi}=0$ for all $\xi$, so $D_1 = 0$. For $N=2$
\be
d_{\xi} = \frac{\pi}{2} \frac{|\xi|}{1 + |\xi|^2} \,,
\ee
and this is maximal on the equator $|\xi| = 1$. So 
\be
D_2 = \frac{\pi}{4} \simeq 0.785 \,.
\ee

For higher $N$ it is not always easy to find $D_N$ by elementary calculus.
A reasonable conjecture is that it occurs in a direction with a 
simple meaning for the relevant platonic solid. Let V, E and F denote 
the sets of directions $\xi$ corresponding to the vertices,
edges and faces of the solid. Let $n_{\rm V}$,
$n_{\rm E}$ and $n_{\rm F}$ denote the number of these elements. By
Euler's theorem, $n_{\rm V}-n_{\rm E}+n_{\rm F}=2$. Because $|\eta|_{\xi}$
has no branch crossing as $\xi$ varies, $d_{\xi}$ is a
smooth Morse function on the 2-sphere. Denote the number of its
local maxima, saddle points and minima by $n_{\rm max}$, $n_{\rm saddle}$ 
and $n_{\rm min}$. By the Morse relations, $n_{\rm max}-n_{\rm saddle}+ 
n_{\rm min}=2$. For a simple function with the symmetry of
the platonic solid, two possible arrangements of the stationary points are
therefore that the maxima occur at V(or F), the saddle points at E and the
minima at F(or V). Exchanging V and F means replacing the solid by its dual.

We will assume that for the spectral curves above, $d_{\xi}$ is such a
simple Morse function. In each case the minimum value of $d_{\xi}$ is zero,
the minima occuring at the roots of the Klein polynomial 
${\cal K}(\xi)$ appearing in the spectral curve equation. For $N=4$, 
${\cal K}(\xi) = \xi^8 + 14\xi^4 + 1$ is the vertex polynomial of the 
cube, so the minima are at the vertices, and the maxima at the face 
centres of the cube. For $N=5$, ${\cal K}(\xi)$ is the same, so the 
conclusion is the same (although we can now say the minima are on the 
faces of an octahedron and the maxima at the vertices). For $N=7$, 
${\cal K}(\xi)=\xi(\xi^{10} + 11\xi^5 - 1)$ is the face polynomial of the
dodecahedron (with one root at infinity), so the maxima of $d_{\xi}$ 
are at the vertices of the dodecahedron. 

The case $N=3$ is more tricky. While the monopole itself has
tetrahedral symmetry, and is not invariant under an inversion, the
function $d_{\xi}$ has this inversion symmetry, because if a line in $\R^3$
belongs to the spectral curve, then so does the line obtained by 
reversing its direction. Inversion sends $\xi$ to its antipode
$-1/\bar\xi$. The distances $d$ for this pair are the same. The
function $d_{\xi}$ therefore has not just tetrahedral symmetry, but cubic
symmetry. For $N=3$, ${\cal K}(\xi)=\xi(\xi^4 - 1)$ is the edge 
polynomial of the tetrahedron, so the edge centres are where $d_{\xi}$ has 
its minima. However, these points are also the face centres of 
the cube whose vertices combine those of the tetrahedron and its
dual. The maxima of $d_{\xi}$ occur on the vertices of this cube, 
that is, on both the faces and vertices of the initial tetrahedron. 
(The 12 saddle points are at the edge centres of the cube, and are 
distributed three per face of the tetrahedron.) Having identified 
the locations of the maxima of $d_{\xi}$ for each of the
platonic monopoles we can now calculate their spectral radii. 

For $N=3$, one vertex of the cube just described is at 
\be
\xi = \frac{1+i}{2} (\sqrt{3}-1) \,,
\ee
this being a root of the Klein polynomial $\xi^8 + 14\xi^4 + 1$.
Solving for $|\eta|_{\xi}$ from the $N=3$ spectral curve gives
\be
d_{\xi} = \frac{\Gamma \left(\frac{1}{6}\right)\Gamma \left(\frac{1}{3}\right)}
{2^{4/3}(3\pi)^{1/2}} \, \frac{|\xi(\xi^4 - 1)|^{1/3}}
{1 + |\xi|^2} \,,
\label{dxi3}
\ee
and evaluating this with the above value of $\xi$, we find 
\be
D_3 = \frac{\Gamma \left(\frac{1}{6}\right)\Gamma \left(\frac{1}{3}\right)}
{2^{4/3}(3\pi)^{1/2}} \, \frac{2^{1/6}}{3^{1/2}} \simeq 1.249 \,.
\ee
It is easy to verify directly that this is the maximum value of $d_{\xi}$.
Firstly, for given $|\xi|$, $\xi^4$ should be negative,
which determines the possible arguments of $\xi$, and then the value 
of $|\xi|$ which maximises the remaining real expression can be found 
by differentiation.

For $N=4$ 
\be
d_{\xi} = \left(\frac{3\Gamma\left(\quart\right)^8}{1024\pi^2}\right)^{1/4}
\, \frac{|\xi^8 + 14 \xi^4 +1|^{1/4}}{1 + |\xi|^2} \,.
\ee
The maxima of $d_{\xi}$ are at the face centres of the cube, the roots
of $\xi(\xi^4 - 1)$, one of which is at $\xi=0$. Hence 
\be
D_4 = \left(\frac{3\Gamma\left(\quart\right)^8}{1024\pi^2}\right)^{1/4}
\simeq 1.725 \,.
\ee
One can verify that $d_{\xi}$ has the same value at other face centres,
e.g. at $\xi=1$, and it has a lower value at the edge centres, e.g. at
$\xi = \sqrt{2} -1$. Again, $D_4$ can be directly verified to be the 
maximum value of $d_{\xi}$. The maximum occurs for $\xi^4$ real and 
non-negative, so we can choose $\xi$ real. Then we can find the 
stationary points of $(\xi^8 + 14\xi^4 + 1)^{1/4}/(1+\xi^2)$ by 
differentiation, and identify the maxima at $\xi=0, \pm 1$ 
(and $\xi = \infty$).
 
For $N=5$ the geometry is the same, but the spectral curve equation
has an extra factor of 4 (and an additional overall factor of
$\eta$). This means that the non-zero roots $\eta$ are multiplied by
$\sqrt{2}$, so $D_5 = \sqrt{2} D_4$. Therefore 
\be
D_5 \simeq \sqrt{2} \times 1.725 \simeq 2.440 \,.
\ee

For $N=7$ we need a vertex of the dodecahedron, a root
of the Klein polynomial $\xi^{20} - 228\xi^{15} + 494\xi^{10} + 228\xi^5 + 1$.
We have found numerically that one of the roots is $\xi = -0.3382612$
with $\xi^5 = -0.00442854$ \cite{num}. Solving the spectral curve 
equation gives
\be
d_{\xi} = \frac{\Gamma \left(\frac{1}{6}\right)
\Gamma\left(\frac{1}{3}\right)}{2\pi^{1/2}} \,
\frac{|\xi(\xi^{10} + 11\xi^5 - 1)|^{1/6}}{1 + |\xi|^2} \,,
\ee
and evaluating this for the numerically obtained $\xi$ gives 
\be
D_7 \simeq 3.176 \,.
\ee
Here we have not shown directly that this is the maximum value of
$d_{\xi}$.

Our combined results are that for $N=1,2,3,4,5,7$, the values of the 
spectral radius $D_N$ are $0, 0.785, 1.249, 1.725, 2.440, 3.176$. For 
comparison the Bolognesi radius is $R_N = N/2$, with values 
$0.5, 1.0, 1.5, 2.0, 2.5, 3.5$. The ratios $D_N/R_N$ are 
$0, 0.785, 0.833, 0.863, 0.976, 0.907$, and these
appear to be approaching 1 as $N$ increases. These calculations
therefore confirm that the platonic monopoles are models
for the Bolognesi magnetic bag at small charge.

Finally, we consider the spectral radii of the sequence of toroidal
monopoles that exist for all $N$. Using the spectral curves obtained
by Hitchin \cite{Hit}, one sees that the largest roots $\eta$ 
for any $\xi$ are
\be
\eta = \pm \frac{N-1}{2} \pi\xi \,,
\ee
so $|\eta|_{\xi} = (N-1)\pi|\xi|/2$. It follows that the maximum value of 
$d_{\xi} = |\eta|_{\xi}/(1 + |\xi|^2)$ is $(N-1)\pi/4$, and this is 
the spectral radius of the charge $N$ toroidal monopole. 
Bolognesi has estimated this using an abelian magnetic disc
approximation, finding a radius $N\pi/4$ (rescaling by 2, as elsewhere in
this section) \cite{Bol1}. These radii agree asymptotically, the
spectral radius being slightly smaller as for the platonic monopoles. 

We see that for any $N$, the toroidal monopole in the magnetic disc 
approximation is larger than the spherical magnetic bag by a factor 
$\pi/2$, which is further evidence that the spherical magnetic bag is the most 
compact monopole. We can also verify that for $N=3,4,5,7$ the spectral 
radius of the toroidal monopole is larger than that of the platonic monopole.

\vspace{7mm}

\section{Conclusions}
\vspace{4mm}

Magnetic bags are clearly an important type of BPS monopole 
solution, even if their existence has not been rigorously
established. The approximately spherical magnetic bag, for each 
large charge $N$, is probably the most compact monopole within 
the whole $4N$-dimensional moduli space of solutions, and we have
found new evidence for this. 

A large class of new monopole solutions has been found, using the same
abelian approximation that gave the magnetic bag solution. These are spherical
clusters of monopoles, and they are all larger than the Bolognesi
bag of the same charge. Some of these clusters have bag structures in their 
interior, and it would be interesting to investigate in more detail 
how the bag joins on to the gas of more isolated monopoles outside the bag. 

We have shown here that the known platonic monopoles with $N=3,4,5,7$ 
form a sequence that look bag-like, and that with increasing $N$ their 
spectral radii approach the asymptotic value of the radius 
predicted for a magnetic bag, $R=N$. There should be further,
exact, bag-like solutions with platonic symmetry. The first of these
has $N=11$. There is then a sequence of rational maps, with icosahedral
symmetry and various degrees from 17 up to 97 and beyond, which have
been used to construct bag-like Skyrmions \cite{BHS}. These could produce
bag-like monopoles too, if one interpreted these rational maps as the
Jarvis maps of monopoles, and implemented the monopole construction 
of Ioannidou and Sutcliffe \cite{IS}. Less clear is the characterisation of 
the Jarvis maps which correspond to the spherical monopole clusters of
variable density that we have discussed. A spherical cluster would 
probably arise from a Jarvis map whose zeros and poles were
distributed randomly but evenly over $S^2$. However, it is difficult to 
construct a monopole of large charge from a rational map of large 
degree, as it is still necessary to solve a partial differential 
equation numerically.

We have interpreted the monopole cluster solutions found here in
astrophysical terms. This was motivated by the apparently good analogy
between a magnetic bag of charge $N$ and a black hole of mass $M$ 
\cite{Bol2}. The bag radius is $R=N$ and (in units where Newton's 
constant and the speed of light are 1) the horizon radius of the black 
hole is $R=2M$.   

For the monopole clusters, we have seen that $\phi$ runs either from $0$ at
the centre to $1$ at infinity, or from $c$ at the centre to $1$ at 
infinity, with $c>0$. Moreover, if $\phi = 0$ at the centre, then there is a 
Bolognesi bag there. These possibilities are analogous to the result
that the Newtonian gravitational potential of an astrophysical 
cluster (in our units) runs from some negative value no less than
$-1/2$ at the centre to $0$ at infinity, and if the central value is 
$-1/2$, then light cannot escape, and there is a central black hole. 
We have seen that the objects we call monopole planets 
have $c>0$ at the centre, so these are like real planets or stars that 
have no black hole inside. The monopole galaxies, on the other hand, 
can have $c=0$. The inner part of such a monopole galaxy is a magnetic 
bag, and this is analogous to a black hole at the centre of a real galaxy.

The conclusion seems to be that clusters of BPS monopoles of large
magnetic charge provide a new, non-relativistic analogy for both 
black holes and less dense astrophysical objects. The analogy
does, of course, have limitations. In particular, the monopole 
solutions considered here are all static, whereas clusters of stars 
could only be approximately stationary if the individual objects 
in the cluster were orbiting the centre. Also, unfortunately, 
monopoles are not physically observable, although they can be studied 
mathematically, and numerical solutions presented visually \cite{book}. 

In summary, the analogy rests on three points. First, both BPS
monopoles and astrophysical objects can be approximately described by
an abelianized scalar potential $\phi$ satisfying a Poisson equation. 
In the monopole case $\phi$ is the abelianized Higgs field; in the 
astrophysical case it is the Newtonian gravitational potential in 
natural units. In both cases, $\phi$ has to be course-grained, to 
smooth out the sharp dips due to unit charge monopoles in 
monopole clusters, or stars within galaxies. Second, the potential 
$\phi$ cannot be less than a critical value: $0$ in the monopole clusters, 
and $-1/2$ in astrophysical objects. In the
monopole case this is not so much because the true Higgs field $\Phi$ 
has minimum magnitude zero at the centre of monopoles, but rather
because $\phi$ determines the local size of a unit charge monopole,
which has to remain finite; in the gravitational case
it is because the Newtonian approximation breaks down where $\phi$
reaches $-1/2$, as light cannot escape from such a region, and the full
machinery of Einstein's equations is needed. Third, the region 
where $\phi$ has its critical value is a region of very compact matter, and 
exhibits an exotic size scaling, with its radius rather than 
volume proportional to the amount of matter. For monopoles, such a 
region is the interior of a Bolognesi magnetic bag, the most compact 
arrangement of magnetic charge. The radius of the bag is equal to 
the bag's magnetic charge. For gravitating matter, the region is the horizon 
of a black hole (with the Newtonian picture invalid inside), whose 
radius is twice the black hole's mass.

\vspace{.5cm}

\section*{Acknowledgements}

This research was largely carried out at the Simons Center for Geometry and
Physics, Stony Brook University. I wish to thank Sir Michael Atiyah
and the SCGP staff for their invitation and hospitality. I also thank
Nigel Hitchin and Sergei Cherkis for discussions.

\end{document}